\documentclass[a4paper,12pt]{article}
\usepackage{graphicx}
\usepackage[american]{babel}

\begin{document}
\date{}

\title{Features of pulsed synchronization of a systems with
a tree-dimensional phase space.}

\author{A.P. Kuznetsov $^1$, N.V. Stankevich $^2$ and L.V. Turukina $^1$}

\maketitle\begin{center} \emph{$^{1}$Institute of Radio
Engineering and Electronics, Russian Academy of Sciences, Saratov Branch\\
Zelenaya 38, Saratov, 410019, Russian Federation\\
$^{2}$ Saratov state University, Astrachanskaya, 83, Saratov,
410012, Russian Federation}\end{center}

\begin{abstract}

Features of synchronization picture in the system with the limit
cycle embedded in a three-dimensional phase space are considered.
By the example of R\"{o}ssler system and Dmitriev - Kislov
generator under the action of a periodic sequence of $\delta$ -
function it is shown, that synchronization picture significantly
depends on the direction of pulse action. Features of
synchronization tons appeared in these models are observed.

\end{abstract}

\section{Introduction}

Synchronization is a fundamental nonlinear phenomenon. It is
presently the subject of intensive interest in many areas of a
science \cite{L1}-\cite{L8}. The classical case of synchronization
consists in an external periodic (usually harmonic) signal acting
upon an autooscillating system with a stable limit cycle
\cite{L3}, \cite{L6}, \cite {L9} - \cite{L12}. In this case,
frequency locking and quasiperiodic regimes are possible inside
and outside the Arnold tongues, respectively, on the external
signal frequency - amplitude plane. In phase space, these regimes
are represented either by the stable torus or by the stable and
saddle limit cycles appearing on this torus. However, there is a
problems when external action has the form of short pulses. This
external action can be represented as periodic sequence of
$\delta-$ functions. Analysis of these problems is important both
for the applications in many areas of science and for the theory
of oscillations and nonlinear dynamics, where it is interesting to
reveal possible specific features of the synchronization picture.
In series of the works \cite{L13} - \cite{L19} this situations
were studied within the framework of model system with a circle
limit cycle under the action of the periodic sequence of $\delta$
- function. For this model approximated one-dimensional circle map
for the phase is constructed. Authors was shown that
synchronization picture both in a model system with a circle limit
cycle and in a one-dimensional map had a number of features. This
synchronization picture essentially different from the classical
as in sine circle map \cite{L6}, \cite{L20}. These features have a
certain degree of universality, since they refer to the
investigation of a system in the vicinity of the Andronov - Hopf
bifurcation. For example, in the papers \cite{L21}, \cite{L22}
similar results was obtained for the Van der Pol and Van der Pol -
Duffing oscillators under the action of the periodic sequence of
$\delta$ - function. At the same time, in systems with
two-dimensional phase space the external pulses can be directed
along any direction in the plane, where the phase portrait of
these systems represents a circle or the object close to circle.
There are more complicated situation in the case then system with
limit cycle embedded in tree-dimensional phase space
\cite{L23},\cite{L24}. Let there is a plane in which limit cycle
is the circle (which is usually the case near the Andronov - Hopf
bifurcation). Then, external pulse acts in any direction in this
plane lead to described above results. Different situation takes
place when the external pulse acts in the direction perpendicular
to this plane. Below, this situation is analyzed in application to
the standard model systems in the nonlinear dynamics. There are
R\"{o}ssler system, which is a artificial system, and Dmitriev -
Kislov generator, which is described real electronic circuit.

\section{Pulsed synchronization of a R\"{o}ssler system}

Let us first consider features of pulsed synchronization by the
example of the R\"{o}ssler system. We start with the case where
the external action (a periodic sequences of $\delta$-functions)
acts in the plane in which the limit cycle of an autonomous
R\"{o}ssler system is situated. Then, the nonautonomous system can
be described as
\begin{equation}
\label{eq11}
  \begin{array}{r c l}
    \dot {x} & = & -y-z+A\sum\delta(t-nT), \\
    \dot {y} & = & x+py, \\
    \dot {z} & = & q+(x-r)z,
  \end{array}
\end{equation}
where $x$, $y$, $z$ are the dynamical variables; $p$, $q$, $r$ are
the system parameters; and $A$ and $T$ are the amplitude and
period of the external action, respectively. Let us select $p$,
$q$, $r$ so as autonomous R\"{o}ssler system has a single limit
cycle of period 1, which is predominantly situated in the ($x$,
$y$) plane and only a small part projects from this plane in the
$z$ axis direction (fig.1).
\begin{figure}[h!]% Fig.1.
\centering
\includegraphics[scale=0.35]{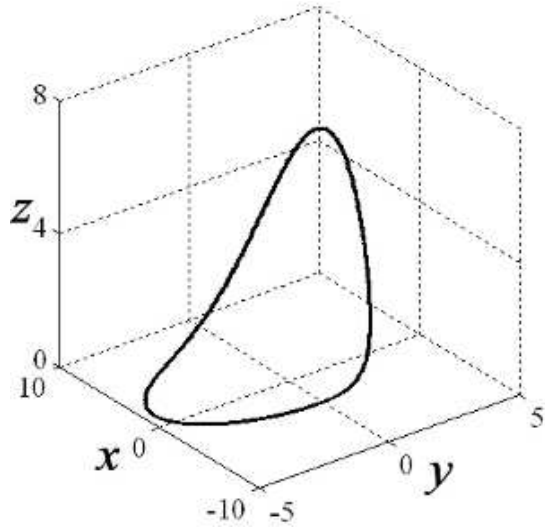}
\caption{Attractor for the autonomous R\"{o}ssler system.
Attractor is plotted for $p=0.2$, $q=0.1$ and $r=1.5$.}
\end{figure}
Depend upon the parameters values altitude of the cycle
"projection" in the $z$ axis direction can be different. As we
shall show further, the synchronization picture in nonautonomous
system depends of these "projection" altitude too.

Let us $p=0.2$, $q=0.1$ and $r=1.5$. The altitude of the cycle
"projection" in the $z$ axis direction in this case is $\Delta z
\approx 3.5$. The period of oscillations in the autonomous system
can be numerically evaluated as $T=5.4368309...$ Figure 2 shows
the chart of dynamical regimes in system (\ref{eq11}) and its
fragments constructed using the results of computer simulation on
the $A - T$ parameters plane.
\begin{figure}[h!]% Fig.2.
\centering
\includegraphics[scale=0.55]{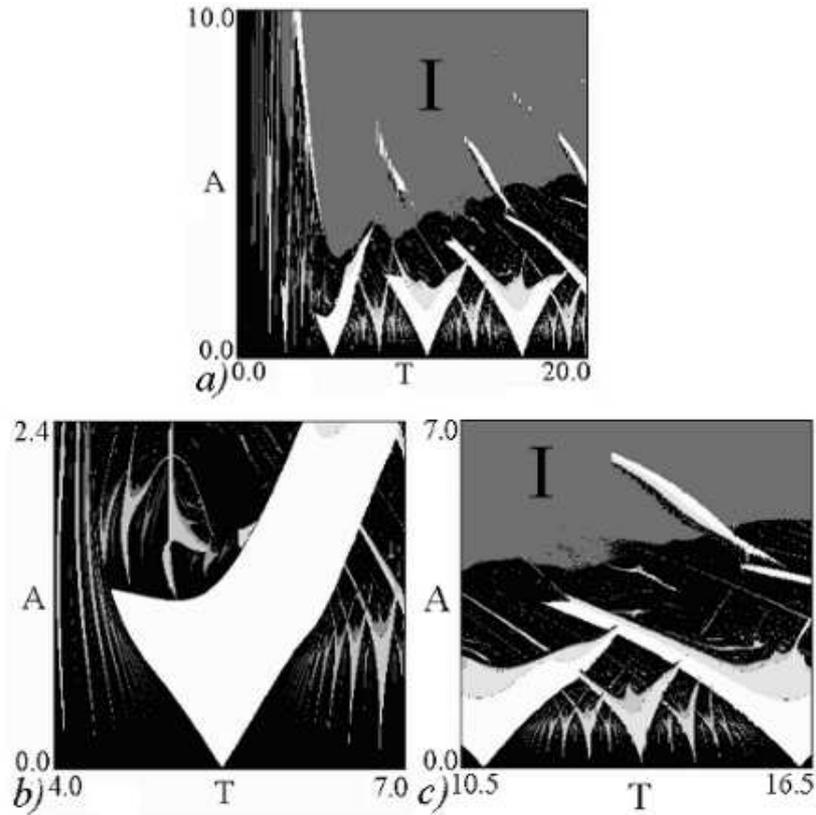}
\caption{The chart of dynamical regimes and its fragments on the
$A - T$ parameters plane, constructed for nonautonomous
R\"{o}ssler system (\ref{eq11}) with the parameters $p=0.2$,
$q=0.1$ and $r=1.5$. Region I corresponds to trajectories going to
infinity.}
\end{figure}
In this and analogous charts to follow, the region painted white
corresponds to the regime of period 1, bright-gray color
corresponds to the regime of period 2 and above, black color
refers to chaos and quasiperiodic regimes and dark-gray color
indicate the region where the trajectory goes to infinity (the
character of each regime is determined in the corresponding
Poincare section). The chart in Fig.2 exhibits a quite traditional
pattern of the Arnold tongs. The increased fragments of this chart
reveal the following features. For the large periods of the
external action (in comparison to the period of autonomous system
$T=5.43...$), the pattern of synchronization is close to classical
that corresponds to the simplest one-dimensional sine circle map
(fig. 2c) \cite{L6}. The internal structure of synchronization
tongs is traditional. One can see the period-doubling bifurcation
and characteristic configurations " crossroad area " \cite{L20}.
On the other hand, for external actions with a period close to the
eigenvalue, the main synchronization tong (albeit surrounded by a
systems of tongs having the classical shape) exhibit a different
internal structure (fig. 2b). This structure reveals a clear line
of the secondary Andronov - Hopf bifurcation and the corresponding
system of secondary synchronization tongs.

Figure 3 shows the nonautonomous attractors of R\"{o}ssler system
(\ref{eq11}).
\begin{figure}[h!]% Fig.3.
\centering
\includegraphics[scale=0.5]{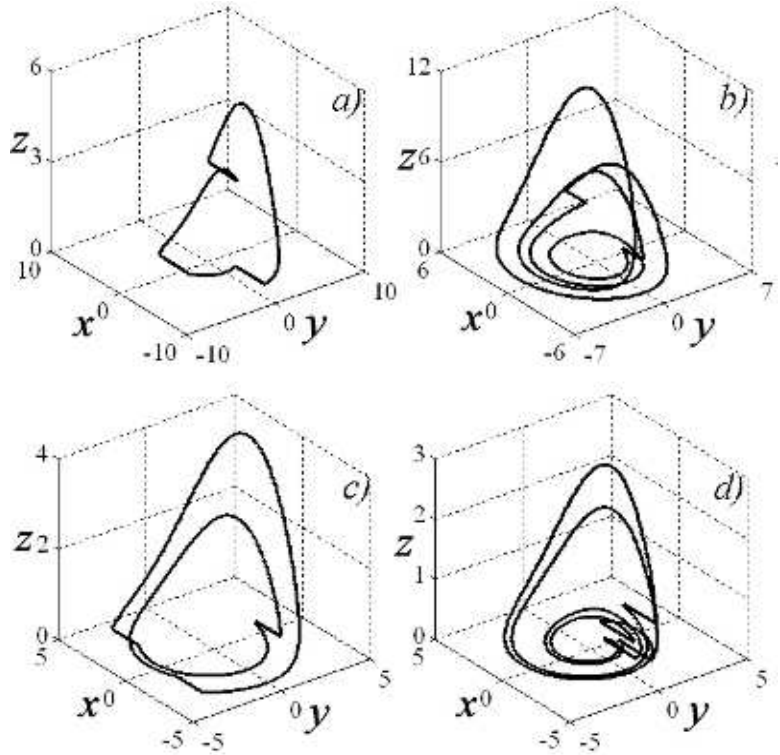}
\caption{The portraits of nonautonomous attractors of R\"{o}ssler
system (\ref{eq11}). Attractors are plotted for the next pulsed
amplitude and period: a) $A=4.4$ and $T=1.8$; b) $A=1.8$ and
$T=10.9$; c) $A=2$ and $T=3.7$; d) $A=2.5$ and $T=7.1$.}
\end{figure}
Figures 3a and 3c shows various configurations of the period 3
attractors, and figures 3b and 3d shows the period 2 and 4
attractors accordingly. All figures concern to a case when the
amplitude of external action is rather essential. If the amplitude
is small, nonautonomous attractors will be distinguished not
strongly from an autonomous case. The external pulse not strongly
changed a trajectory of autonomous system. Thus, the period of
noautonomous system increased, but the attractor shape will change
a little.

Now we change parameters of the system so, that the altitude of
the cycle "projection" in the $z$ axis direction increased. Let us
$p=0.3$, $q=0.1$, $r=2.0$. In this case $\Delta z \approx 6.7$.
The chart of dynamical regimes and its fragments on the $A - T$
parameters plane, constructed for nonautonomous R\"{o}ssler system
(\ref{eq11}) for this case are presented in Fig. 4.
\begin{figure}[h!]% Fig.4.
\centering
\includegraphics[scale=0.55]{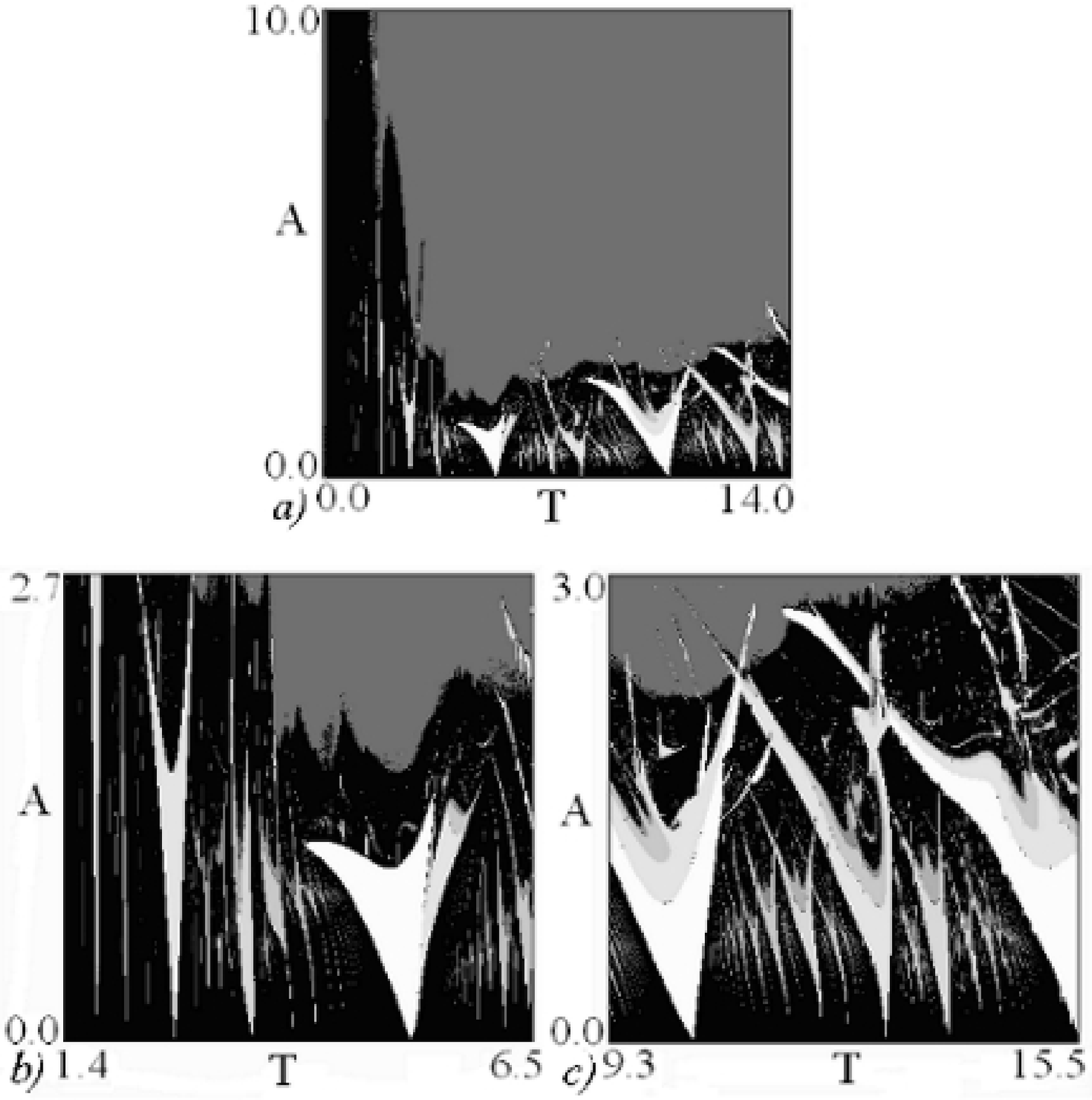}
\caption{The chart of dynamical regimes and its fragments on the
$A - T$ parameters plane, constructed for nonautonomous Ressler
system (\ref{eq11}) with the parameters $p=0.3$, $q=0.1$ and
$r=2.0$.}
\end{figure}
One can see, that in this case, as well as in previous,
synchronization tongs are observed. However, the region in which
the trajectories going to infinity has increased. And the periodic
and quasiperiodic regimes are realized at smaller values of
amplitude of external action(fig. 4a). Also Arnold tongs were a
little transformed (fig. 4 b, c). They became narrower, their
symmetry is broken, and they are a little bit inclined in the left
side.

Now let us consider how the synchronization picture in a
nonautonomous R\"{o}ssler system changes when the pulses act along
the $z$ axis, that is perpendicularly to the plane to which the
limit cycle predominantly belong:
\begin{equation}
\label{eq12}
  \begin{array}{r c l}
    \dot {x} & = & -y-z, \\
    \dot {y} & = & x+py, \\
    \dot {z} & = & q+(x-r)z+A\sum\delta(t-nT),
  \end{array}
\end{equation}
Again we shall start with a case when autonomous R\"{o}ssler
system has a single limit cycle of period 1 and altitude of the
cycle "projection" in the $z$ axis direction is $\Delta z \approx
3.5$, i.e. $p=0.2$, $q=0.1$ and $r=1.5$. Figure 5 shows the chart
od dynamical regimes and its magnified fragments for this case.
\begin{figure}[h!]% Fig.5.
\centering
\includegraphics[scale=0.55]{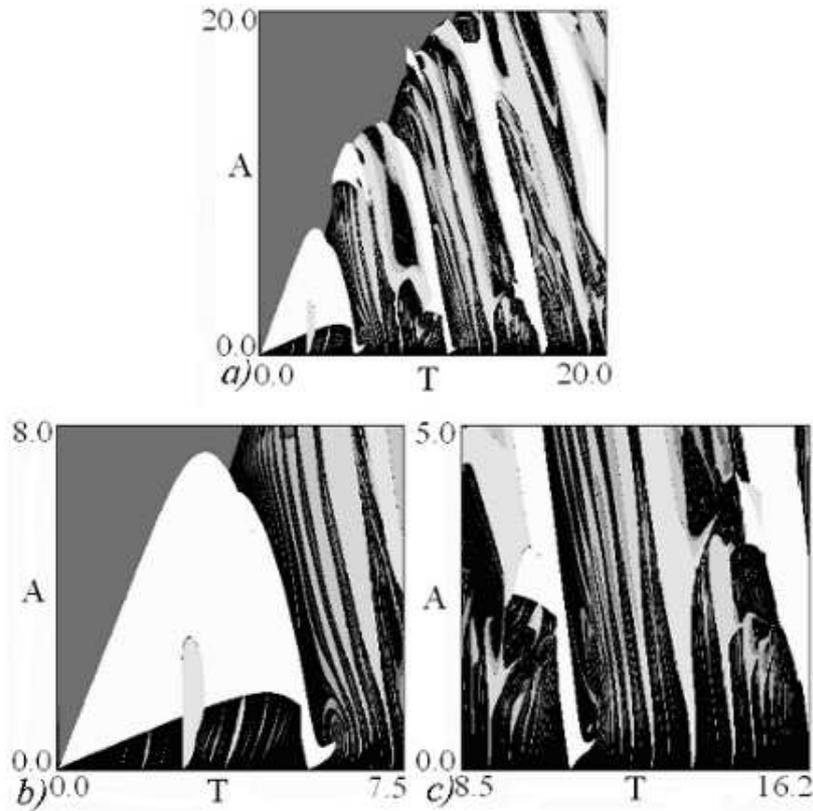}
\caption{The chart of dynamical regimes and its fragments on the
$A - T$ parameters plane, constructed for nonautonomous
R\"{o}ssler system (\ref{eq12}) with the parameters $p=0.2$,
$q=0.1$ and $r=1.5$.}
\end{figure}
A comparison between Figs. 2 and 5 shows substantial changes in
the pattern of synchronization.  In particular, the map of system
(\ref{eq12}) in Fig. 2 exhibits, in addition to the system of
synchronization tongs and the region of quasi-periodic regimes
(situated in the bottom part of the chart), rather large regions
of the regimes of period 1 and 2, which correspond to higher
values of the amplitude of external action. The lower boundary of
the regimes of period 1 represents the line of the Neimark -
Sucker bifurcation. The Arnold tongs touching this line have sharp
vertices both at the base and the top (Fig. 5b). There is a small
system of tongs forming open rings in the right-hand part of the
chart. The other tongs on the right from the main one appear as
bands. For large amplitudes these tongs transform into large
regions corresponding to the regimes of various periods (Figs. 5a
and 5c). Note that in the case of small periods of external action
the threshold from which the trajectories go to infinity shifts to
zero. This behavior also differs from that observed in Fig. 1. The
nonautonomous periods 1, 2, 3 and 5 attractors of R\"{o}ssler
system (\ref{eq12}) are presented in Figure 6 which should be
compared to Fig. 3.
\begin{figure}[h!]% Fig.6.
\centering
\includegraphics[scale=0.5]{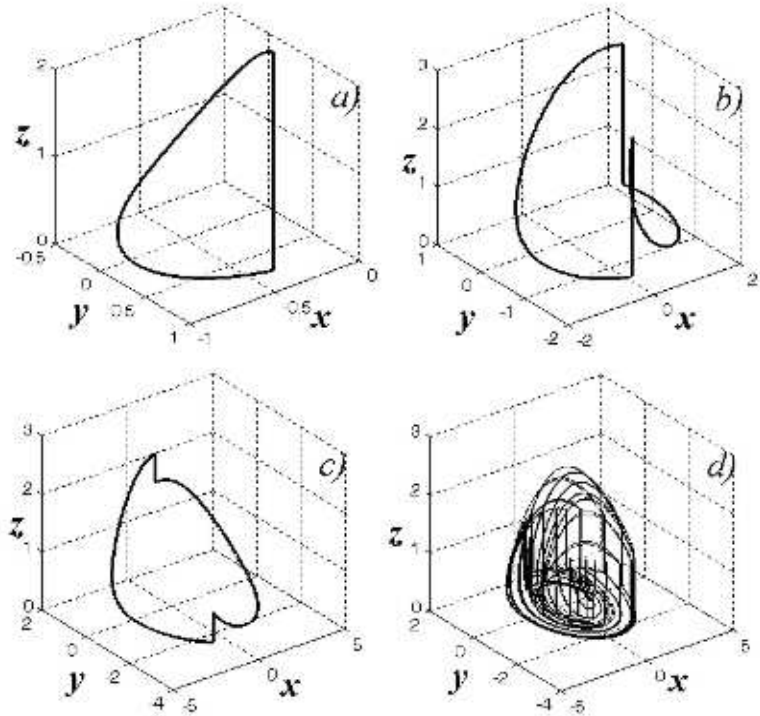}
\caption{The portraits of nonautonomous attractors of R\"{o}ssler
system (\ref{eq12}). Attractors are plotted for the next pulsed
amplitude and period: a) $A=5$ and $T=2.85$; b) $A=2.4$ and
$T=2.85$; c) $A=0.5$ and $T=2.85$; d) $A=1$ and $T=4.5$.}
\end{figure}

Now let us changed parameters of the system so, that the altitude
of the cycle "projection" in the $z$ axis direction increased and
$\Delta z \approx 6.7$. Figure 7 shows the corresponding chart of
dynamical regimes and its magnified fragments.

A comparison between Figs. 4 and 7 shows that the region in which
the trajectories go to infinity has increased and region of the of
periodic and quasi-periodic regimes has decreased. This is
especially well visible at large values of periods of external
action. Besides synchronization tongs become narrower and are a
inclined in the left side. They internal structure becomes
complicated. The region of small values of period of external
action ($T<6.0$) is exception. Here the region of periodic and
quasi-periodic regimes increased. And there are transition to
chaos through period-doubling bifurcation inside synchronization
tongs.
\begin{figure}[h!]% Fig.7.
\centering
\includegraphics[scale=0.55]{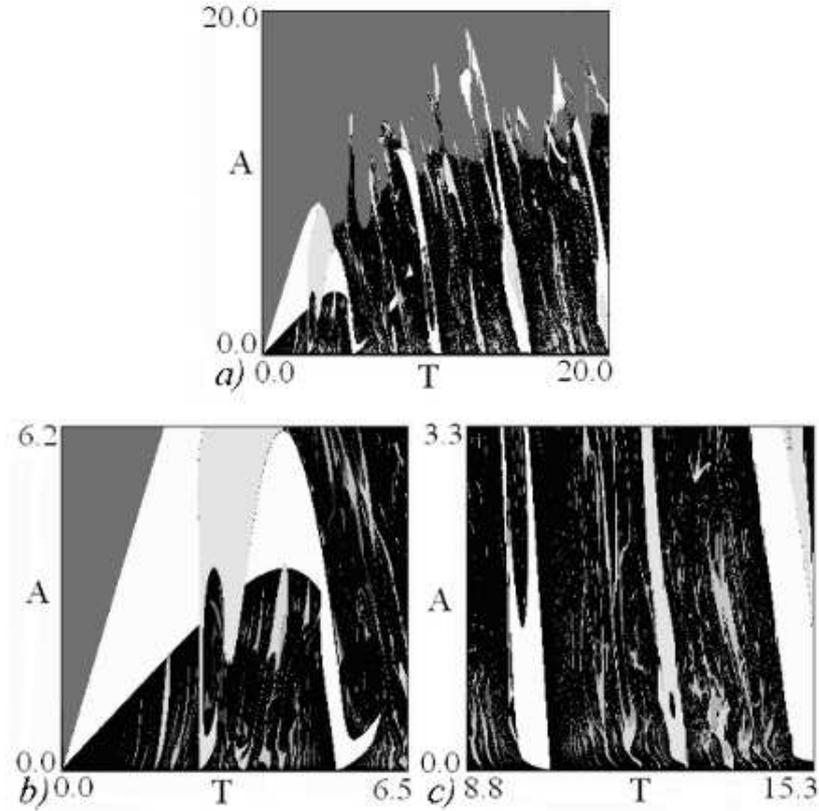}
\caption{The chart of dynamical regimes and its fragments on the
$A - T$ parameters plane, constructed for nonautonomous
R\"{o}ssler system (\ref{eq12}) with the parameters $p=0.3$,
$q=0.1$ and $r=2.0$.}
\end{figure}

\section{Pulsed synchronization of a Dmitriev - Kislov generator}

Now let us consider features of pulsed synchronization by the
example of the Dmitriev - Kislov generator \cite{L20}, \cite{L24}:
\begin{equation}
\label{eq1}
  \begin{array}{r c l}
    x+D\dot {x} & = & Mz\exp(-z^{2}), \\
    \dot {y} & = & x-z, \\
    \dot {z} & = & y-\frac{z}{Q},
  \end{array}
\end{equation}
where $x$, $y$, $z$ are the dynamical variables, $D$, $M$, $Q$ are
the system parameters. Dynamics of this system is studied
adequately. Thus we selected the system's (\ref{eq1}) parameters
so that in phase space there is the period-1 stable limit cycle.
Let us fix $D=6$, $M=5.5$ and $Q=10$. The tree-dimensional
attractor for system (\ref{eq1}) and its projections are presented
in Fig. 8.
\begin{figure}[h!]% Fig.8.
\centering
\includegraphics[scale=0.6]{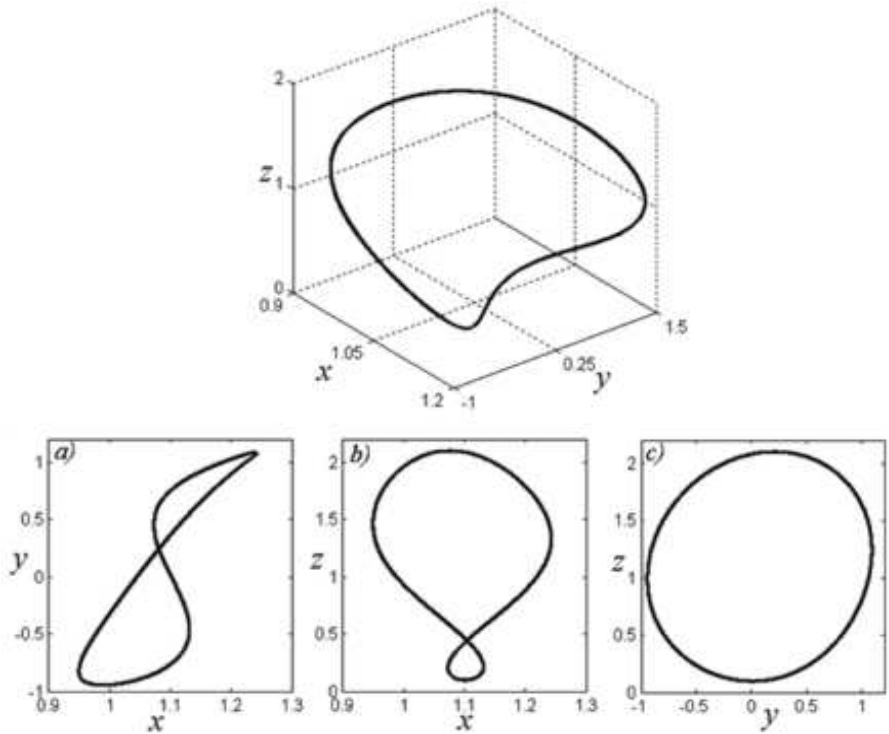}
\caption{Attractor and its projections for the Dmitriev - Kislov
generator (\ref{eq1}). Attractor and projections are plotted for
$D=6$, $M=5.5$ and $Q=10$.}
\end{figure}
One can see that projection of the attractor to the ($y$, $z$)
plane is circle (Fig. 8 b). Two other projections is different
form the circle, essentially, projection to the ($x$, $y$) plane.
It is looks like the figure-of-eight (Fig. 8 a).

Let us add into Dmitriev - Kislov generator external pulses in the
form $A\sum\delta(t-nT)$. Here $A$ is the pulsed amplitude and $T$
is the pulsed period. Note, that external pulses add each
equations by turns. Thus, let us consider next situations. First
case, when pulses act along the $x$ axis, that is external action
add to the fist equation. Second case, when pulses act along the
$y$ axis, that is external action add to the second equation.
Third case, when pulses act along the $z$ axis, that is external
action add to the third equation. Figure 9 shows the charts of
dynamical regimes on the pulsed amplitude - period plane.
\begin{figure}[h!]% Fig.39.
\centering
\includegraphics[scale=0.55] {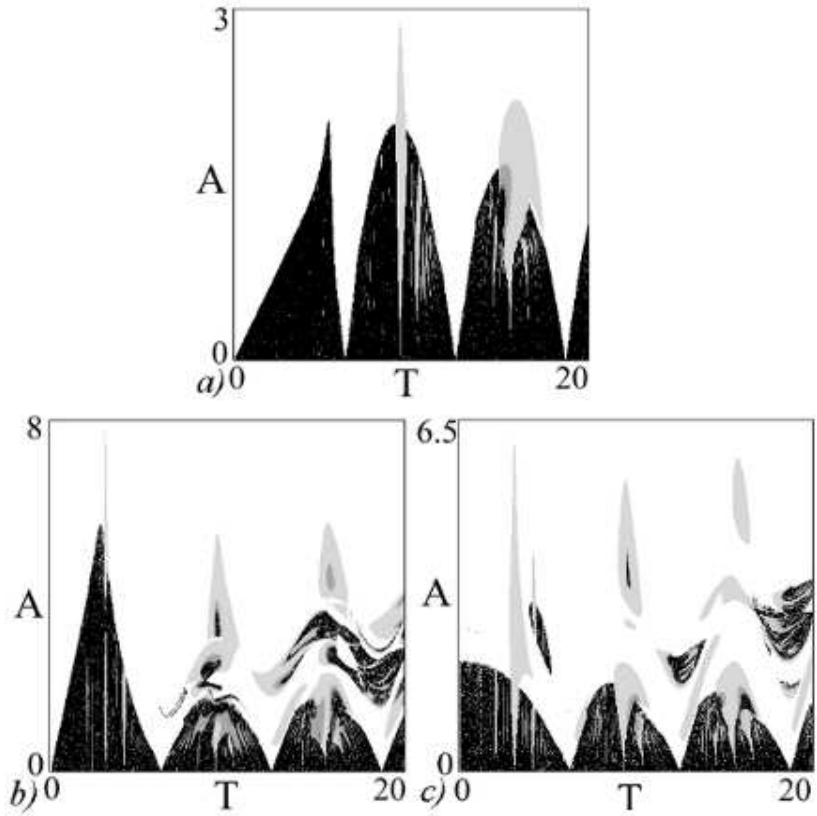}
\caption{Charts of dynamical regimes of the nonautonomous Dmitriev
- Kislov generator on the $A$ - $T$ plane. Charts are plotted for
the cases, when pulses act along the $x$ axis (fig.a), the $y$
axis (fig.b) and the $z$ axis (fig.c). Systems parameters are
fixed at $D=6$, $M=5.5$ and $Q=10$.}
\end{figure}
These charts are plotted for all cases referred above. Chart of
dynamical regimes is the plane of parameters, at which regions of
periodic behavior with different periods are shown with different
colors. In this and analogous charts to follow, the region painted
white corresponds to the regime of period 1, bright-gray color
corresponds to the regime of period 2 and above, black color
refers to chaos and quasiperiodic regimes. As can be seen from
Figs. 2 and 3, the case when pulses act along the $y$ axis is most
interesting. In this case synchronization picture differ both from
the classical typical for the sine circle map and from the
synchronization picture typical for system with a circle limit
cycle under the action of the periodic sequence of $\delta$ -
function. We can explains this fact as follows. In the case then
external action add to the second equation of the system
(\ref{eq1}) (to the dynamical variable $y$) pulses will be crossed
both loops of the figure-of-eight (Fig. 8 a). In the other cases
pulses are crossed either circle or one loop of the
figure-of-eight. As a result, synchronization picture in the last
two cases is the same like in the model with circle limit cycle
under the action of the periodic sequence of $\delta$ - functions
(Figs. 3a and 3b). Thus, let us discuss synchronization in
nonautonomous Dmitriev - Kislov generator in the case then
external pulses act along the $y$ axis more detailed.

The fragments of the chart of dynamical regimes of nonautonomous
attractors for the Dmitriev - Kislov generator are presented in
the Figs. 10 and 11.
\begin{figure}[h!]% Fig.10.
\centering
\includegraphics[scale=0.7] {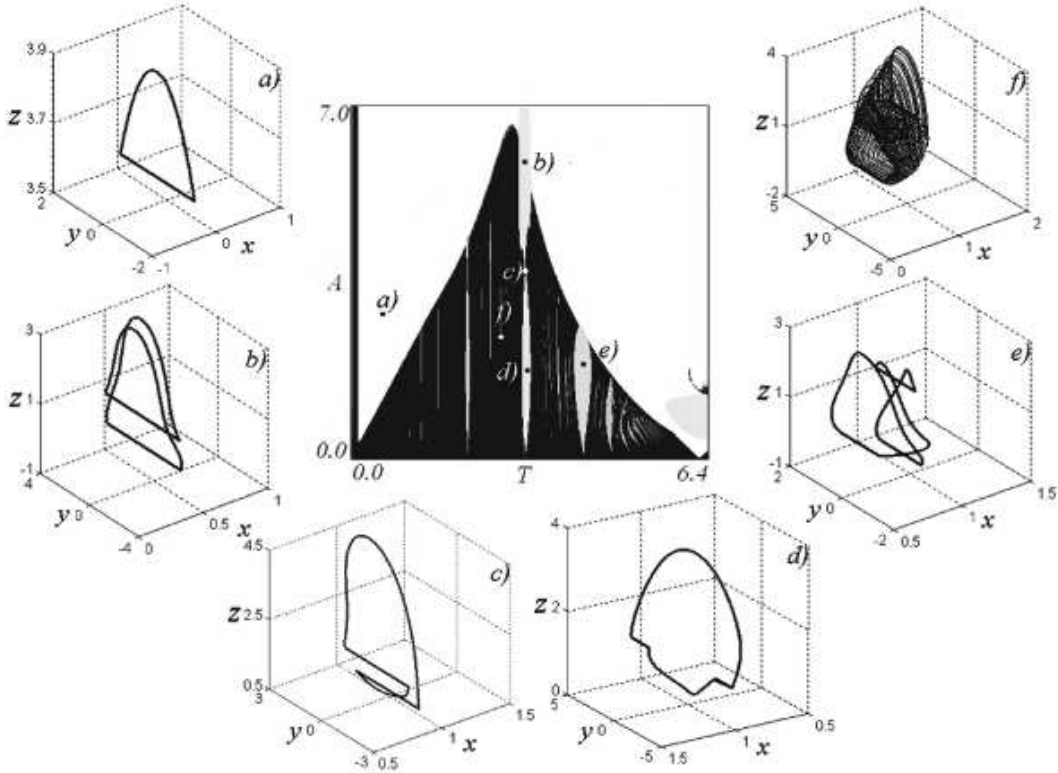}
\caption{Fragment of the chart of dynamical regimes for the
nonautonomous Dmitriev - Kislov generator (\ref{eq1}) from the
Fig.9b. The portraits of nonautonomous attractors are shown in
this figure too. Attractors are plotted for the next pulsed
amplitude and period: a) $A=3$ and $T=0.8$; b) $A=6$ and $T=3.3$;
c) $A=3.7$ and $T=3.13$; d) $A=2$ and $T=3.14$; e) $A=1.5$ and
$T=4.1$; f) $A=3$ and $T=3$.}
\end{figure}
\begin{figure}[h!]% Fig.11.
   \leavevmode
\centering
\includegraphics[scale=0.6] {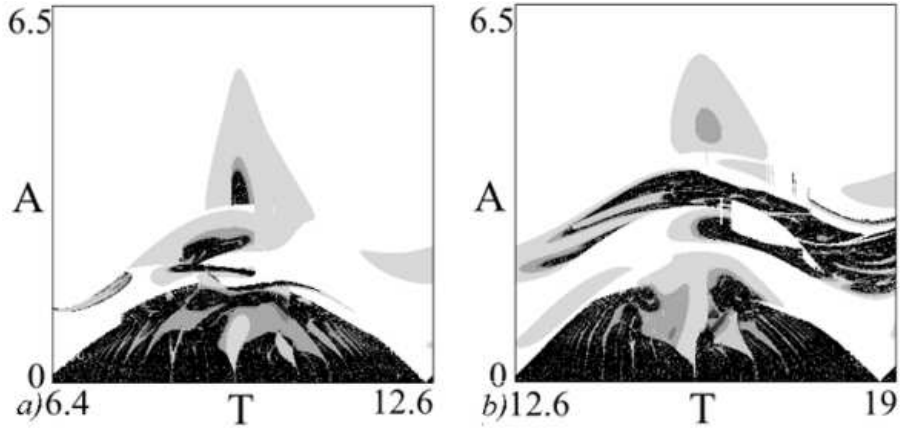}
\caption{Fragments of the chart of dynamical regimes for the
nonautonomous Dmitriev - Kislov generator (\ref{eq1}) from the
Fig. 9 b.}
\end{figure}
In the charts one can see synchronization tongs corresponded to a
periods multiple to the eigenvalue (the natural period of the
system (\ref{eq1}) $T\approx6.22$). Region of a quasiperiodic
regimes are surrounded these tongs. Higher-order synchronization
tongs are observed in this region. Synchronization tongs
corresponded to subharmonic resonance are located in the region of
lesser periods of external action (comparison to the natural
period). These synchronization tong looks like strongly stretch up
"drops". But note that, synchronization tong with rotation number
$\frac{1}{2}$ (period 2) differ from the others synchronization
tongs in this region. This tong is touched to $A=0$ axis. First,
it is converged in the middle part by forming "bridge". Then it is
widen in the top part. Besides, in the top part of the chart this
synchronization tong comes to the period 1 region (Fig.4). Thus,
on the one hand it is synchronization tong with rotation number
$\frac{1}{2}$. On the other hand it is the period 2 region, which
is originated from the regime of period 1 being at large values of
the pulsed amplitude as a result of period-doubling bifurcation.
Nonautonomous attractors plotted for the Dmitriev - Kislov
generator is verified this conclusion (Fig.4a-f). Thus, attractor
plotted in the upper part of the period 2 synchronization region
(Fig. 10 b) looks like period 2 attractor originated from the
period 1 attractor (Fig. 10 4a) as a result of period-doubling
bifurcation. But, attractor plotted in the lower part of the
period 2 synchronization region (Fig. 10 c) looks different. It is
corresponded to regime of synchronization with rotation number
$\frac{1}{2}$. Transition between these to types of the
nonautonomous attractors observed under decrease of pulsed
amplitude. This transition consist in the next. One attractor loop
(see Fig. 10 b) is decreased throughout the height. While in the
"bridge" it is situated in the same plane that action (Fig. 10 c).
After that loop is lower oneself under this plane (Fig. 10 d). It
is important note, that this process is accompanied by decrease of
pulsed amplitude. These features are not observed for all others
synchronization tongs in this region (Fig. 10).

Synchronization tongs of different periods and quasiperiodic
regimes are observed on the right of main tong (Fig. 11) However,
these tongs are smaller by the amplitude and had more complicated
structure. Inside these tongs there are  regions of different
period and chaos, i.e. transition to chaos through period-doubling
cascade. Moreover, there are other regions of the complex dynamics
in the top part of the charts presented in Fig. 11. Thus, one can
conclude that in the nonautonomous system (\ref{eq1}) there are
periodic and quasiperiodic regimes at the small values of the
pulsed amplitude and periodic regimes and chaos at the large
values.

\section{Conclusion.}

Thus by the example of R\"{o}ssler system and Dmitriev - Kislov
generator the problem of synchronization by the periodic sequence
of $\delta$ - function in the system with a three-dimensional
phase space are considered. Different methods of the adding of
external action to the system are studied. It is showed that the
synchronization in the system with limit cycle embedded in a
tree-dimensional phase space significantly depends on the
direction of pulse action and on the attractor's form. If external
pulses act in the plane, where the projection of phase portrait of
autonomous system represents a circle (or the object close to a
circle), then synchronization picture in nonautonomous system will
be closed to the synchronization picture typical for
two-dimensional system with a circle limit cycle under the acting
of the periodic sequence of $\delta$ - function. However, if
projection of phase portrait essentially differ from the circle or
if the external pulses act in the direction perpendicular to the
plane of the limit cycle, then synchronization picture will be
different. In the present paper it is showed that this
synchronization picture will be differed both from the classical
typical for the sinus-circle map and from the synchronization
picture typical for two-dimensional system with a circle limit
cycle under the action of the periodic sequence of $\delta$ -
function.

\textit{This study was supported by the Russian Foundation for
Basic Research (project no. 06-02-16773) and CRDF BRHE (grant
REC-006 Y2-P-06-13.)}

\end{document}